\documentclass[11pt,a4paper]{article}
\usepackage[qm]{qcircuit}

\usepackage{mathtools}
\usepackage{authblk} % author
\usepackage{algpseudocode,algorithmicx,algorithm}

\usepackage{mathrsfs}
\usepackage{latexsym,bm}
\usepackage{amsmath,amsfonts,amssymb,amsthm}
\usepackage{extarrows}

\usepackage{graphicx,subfigure,epstopdf,float}
\usepackage{enumerate,cases,multirow}
\usepackage{makecell}
\usepackage{caption}

\usepackage{longtable,colortbl,arydshln,threeparttable}
\definecolor{mygray}{gray}{.9}

\usepackage{indentfirst}
\usepackage[top=25mm,bottom=20mm,left=25mm,right=20mm]{geometry}
\usepackage{cite}

\usepackage{listings}

\usepackage{makeidx}        % generate an index, automatically
\usepackage{booktabs}
\usepackage[bookmarks,bookmarksnumbered,colorlinks,citecolor=red,linkcolor=red,hyperindex,linktocpage=true]{hyperref}

\newcommand{\ket}[1]{| #1 \rangle} % |u>
\newcommand{\bra}[1]{\langle #1 |} % <u|

\newcommand{\bb}{\boldsymbol}

\def \d {\mathrm{d}}
\def \e {\mathrm{e}}
\def \i {\mathrm{i}}

\newcounter{parentalgorithm}

\makeatother

\newtheorem{theorem}{Theorem}[section]
\newtheorem{lemma}{Lemma}[section]

\theoremstyle{remark}

\numberwithin{equation}{section}

\begin{document}

\title{\bfseries Schr\"odingerization for quantum linear systems problems with near-optimal dependence on matrix queries}

\author[1,2,3]{Yin Yang\thanks{yangyinxtu@xtu.edu.cn}}
\author[1,2,3]{Yue Yu\thanks{terenceyuyue@xtu.edu.cn}}
\author[1,2,3]{Long Zhang\thanks{longzhang@smail.xtu.edu.cn}}

\affil[1]{School of Mathematics and Computational Science, Xiangtan University, Xiangtan, Hunan 411105, China}
\affil[2]{Hunan Research Center of the Basic Discipline Fundamental Algorithmic Theory and Novel Computational Methods, Xiangtan, Hunan 411105, China}
\affil[3]{National Center for Applied Mathematics in Hunan, Xiangtan, Hunan 411105, China}

\maketitle

% REQUIRED
\begin{abstract}
We develop a quantum algorithm for linear algebraic equations $ A\bb{x} = \bb{b} $ from the perspective of Schr\"odingerization-form problems, which are characterized by a system of linear convection equations in one higher dimension.
  When $ A $ is positive definite, the solution $ \bb{x} $ can be interpreted as the steady-state solution to a system of linear ordinary differential equations (ODEs). This ODE system can be solved by using the linear combination of Hamiltonian simulation (LCHS) method in \cite{ACL2023LCH2}, which serves as the continuous implementation of the Fourier transform in the Schr\"odingerization method from \cite{JLY22SchrShort, JLY22SchrLong}. Schr\"odingerization transforms linear partial differential equations (PDEs) and ODEs with non-unitary dynamics into Schr\"odinger-type systems via the so-called warped phase transformation that maps the equation into one higher dimension.
  When $ A $ is a general Hermitian matrix, the inverse matrix can still be represented in the LCHS form in \cite{ACL2023LCH2}, but with a kernel function based on the Fourier approach in \cite{Childs2017QLSA}. Although this LCHS form provides the steady-state solution to a system of linear ODEs associated with the least-squares equation, applying Schr\"odingerization to this least-squares system is not appropriate, as it results in a much larger condition number.
  We demonstrate that in both cases, the solution $ \bb{x} $ can be expressed as the LCHS of Schr\"odingerization-form problems. We provide a detailed implementation and error analysis. Furthermore, we incorporate a block preconditioning technique to achieve nearly linear scaling in the condition number, thereby attaining near-optimal query complexity.
\end{abstract}

% REQUIRED

\section{Introduction}

Quantum computing is an emerging computational paradigm that has attracted significant attention, primarily due to the discovery of quantum algorithms capable of offering exponential speedups over the best-known classical methods \cite{Nielsen2010, LR2010QuantumAlgebra, Deutsch1992rapid, Shor1997Prime, HHL2009}. Numerous quantum algorithms for scientific computing have been proposed in recent years. One fundamental task that underlies many areas of science and technology is the development of solvers for a linear system of equations
\begin{equation}\label{Linearsystem}
A\bb{x} = \bb{b},
\end{equation}
where $A$ is an $N\times N$ invertible matrix with $N = 2^{n}$. In the realm of quantum computing, these are known as quantum linear systems algorithms (QLSAs), with the HHL algorithm \cite{HHL2009, Childs2017QLSA, Costa2022QLSA} being a prominent example.

Classical solvers for linear systems typically take time proportional to the number of unknown variables, making them computationally expensive for large systems. This is especially true for systems arising from the numerical discretization of high-dimensional partial differential equations (PDEs). The HHL algorithm, introduced by Harrow, Hassidim, and Lloyd in 2009 \cite{HHL2009}, is the first quantum algorithm proposed for solving linear systems. Its gate complexity is $\mathcal{O}(\log(N) s^2\kappa^2 /\delta )$, where $ \kappa $ is the condition number of the matrix, $ s $ is the sparsity of $ A $, and $ \delta $ is the precision. In contrast, for positive-definite matrices, the Conjugate Gradient (CG) method has a time complexity of $\mathcal{O}(N s\sqrt{\kappa} \log (1/\delta) $, which exhibits polynomial growth with respect to $ N $, making the HHL algorithm exponentially faster for large systems when the condition number $ \kappa $ of the matrix is not excessively large.
To address the limitations of phase estimation in the HHL algorithm, Ambainis \cite{Ambainis2012VTAA} introduced the variable-time amplitude amplification (VTAA) method, which improves the dependence on the condition number, reducing it from $\kappa^2$ to $\kappa \log^3(\kappa)$.
Building on this, Childs et al. \cite{Childs2017QLSA} further improved the algorithm, achieving a nearly linear dependence on the condition number.
To enhance the accuracy dependence, they replaced the Hamiltonian simulation with either the Fourier method or the Chebyshev method, resulting in a $\text{polylog}(1/\delta)$ dependence. This is similar to the dependence observed in classical methods and provides an exponential acceleration in accuracy compared to the HHL algorithm.
However, the VTAA procedure is highly complex, involving multiple rounds of recursive amplitude amplification, which makes it difficult to implement in practice \cite{Costa2022QLSA}.
To address this issue, alternative approaches based on adiabatic quantum computing (AQC) have been developed in recent years.
Dong et al. \cite{An-Lin-2022} proved that by employing an optimally tuned scheduling function, AQC is capable of efficiently solving a quantum linear system problem (QLSP) with a runtime of $ \mathcal{O}(\kappa  \text{poly}\log(\kappa/\delta)) $. Lin et al. \cite{Lin-Tong-2020} proposed a quantum eigenstate filtering algorithm and introduced Zeno eigenstate filtering when applying it to the QLSP. One can also refer to \cite{Costa2022QLSA} for a comprehensive review of the literature along this line, where the optimal scaling with the condition number is achieved with query complexity $\mathcal{O}(\kappa \log (1/\delta))$ by using a discrete quantum adiabatic theorem proved in \cite{DKS1998Adiabatic}, which completely avoids the heavy mechanisms of VTAA or the truncated Dyson-series subroutine from previous methods related to the AQC \cite{Subasi2019AQC}.

In this article, we aim at developing quantum algorithms for solving linear algebraic equations from the perspective of Schr\"odingerization-form problems, which are characterized by a system of linear convection equations in one higher dimension. This approach is straightforward when $ A $ is positive definite. In this case, we can consider the solution $\bb{x}$ as the steady-state solution to the system of linear ordinary differential equations (ODEs) $\bb{u}_t = -A \bb{u} + \bb{b}$, as described in \cite{HJZ2024multiscale}. The resulting linear ODE system is then solved by the Schr\"odingerization method introduced in \cite{JLY22SchrShort, JLY22SchrLong}. Schr\"odingerization transforms linear PDEs and ODEs with non-unitary dynamics into Schr\"odinger-type systems via the so-called warped phase transformation that maps the equation into one higher dimension.
The ODE system can also be resolved by using the linear combination of Hamiltonian simulation (LCHS) method in \cite{ACL2023LCH2}, which can be viewed as the continuous implementation of the Fourier transform in the Schr\"odingerization method.

For a general Hermitian matrix, it is unclear whether a Schr\"odingerization-based algorithm can be developed. However, it is evident that the inverse matrix $A^{-1}$ can still be represented in the LCHS form from \cite{ACL2023LCH2}, but with a kernel function based on the Fourier approach in \cite{Childs2017QLSA}. While this LCHS form provides the steady-state solution for a system of linear ODEs with coefficient matrix $-A^2$ and source term $A\bb{b}$, corresponding to the least-squares equations $A^2\bb{x} = A\bb{b}$, applying Schr\"odingerization to this ODE system is not appropriate, as it results in a significantly larger condition number.
Inspired by the similarities in the LCHS forms and recognizing that the LCHS method for linear ODEs can be recast as the Schr\"odingerization method, we successfully express the solution $\bb{x}$ as the LCHS of Schr\"odingerization-form problems. 
%By applying the Duhamel's principle, we instead reformulate this integral form of the solution as the steady-state solution to a Schr\"odingerization-form problem.
%This highlights the potential of the Schr\"odingerization technique in quantum scientific computation.
Based on this result, we are in a position to develop a quantum algorithm utilizing the discrete Fourier transform as done for the Schr\"odingerization method.

The time complexity of our algorithm still exhibits a quadratic dependence on the condition number. However, by employing the block preconditioning technique proposed in \cite{Low2026quantumlinearsystem}, we reduce the $\kappa$-dependence from quadratic to nearly linear. 

The paper is organized as follows.
In Section \ref{sec:LCHSQLSP}, we introduce an abstract framework that expresses the inverse matrix as a linear combination of Hamiltonian simulations, characterized by kernel functions. We then demonstrate how the LCHS formula can be transformed into an LCHS with a Hamiltonian defined by a system of linear convection equations in one higher dimension, which is referred to as a Schr\"odingerization-form problem in this work.
In Section \ref{sec:implementationSchrQLSP}, we provide detailed implementation procedures along with an error analysis. 
Section \ref{sec:preconditioning} combines our algorithm with the block preconditioning technique in \cite{Low2026quantumlinearsystem} to achieve nearly linear scaling in the condition number.

\section{Schr\"odingerization for quantum linear systems problems} \label{sec:LCHSQLSP}

This section demonstrates that the matrix inverse can be expressed as a linear combination of Hamiltonian simulations, where the Hamiltonians are given by convection equations in one higher dimension. These convection equations are referred to as Schr\"odingerization-form problems.
	
\subsection{LCHS for matrix inverse}
	
We first introduce an abstract framework of solving the quantum linear system problem (QLSP), which represents the QLSP as a linear combination of Hamiltonian simulations $\e^{-\i H t_i}$, where $t_i \in \mathbb{R}$ and $H$ is a Hermitian matrix defined by the coefficient matrix of the linear system.
	
\begin{theorem} \label{thm:abstractLCHSQLSA}
Let $ A $ be an invertible Hermitian matrix, and let $\varphi(k)$ be a function defined on $\mathbb{R}$, with its Fourier transform given by $\hat{\varphi}(s) = \int_{\mathbb{R}}  \varphi(k) \e^{-\i k s} \d k$. Assume that
\[\int_0^\infty \hat{\varphi}(\lambda s) \d s = \frac{1}{\lambda}, \qquad 0\ne \lambda \in [\lambda_{\min}, \lambda_{\max}],\]
where $\lambda_{\min}$ and $\lambda_{\max}$ are the minimum and maximum eigenvalues of $A$, respectively. Then the inverse matrix $A^{-1}$ can be represented as a linear combination of Hamiltonian simulation problems:
\begin{equation}\label{abstractinvA}
A^{-1} =  \int_0^{\infty} \int_{\mathbb{R}}  \varphi(k) \e^{-\i k A s} \d k \d s.
\end{equation}
\end{theorem}
	\begin{proof}
		The result is obtained through the diagonalization of $A = U \Lambda U^\dag$, where $\Lambda = (\lambda_1,\cdots,\lambda_N)$ is a diagonal matrix and $U$ is a unitary matrix.
	\end{proof}
	
According to the Fourier approach in \cite{Childs2017QLSA}, the kernel function can be chosen as
\begin{equation}\label{varphik}
\varphi(k) = \frac{\i}{\sqrt{2\pi}} k \e^{-k^2/2}, \qquad \hat{\varphi}(s)  = s\e^{-s^2/2},
\end{equation}
which does not impose any requirements on the signs of the eigenvalues of $A$.

It is clear that $\bb{x}$ can be interpreted as the steady-state solution to the following system of ODEs:
\begin{equation}\label{underlyingODE}
\frac{\d \bb{u}}{\d t} = - A^2\bb{u} + A \bb{b}, \quad \bb{u}(0) = \bb{0}.
\end{equation}
One can verify that
\[\|\bb{x} - \bb{u}(T_u)\| \le \delta \|\bb{x}\| \quad \text{if} \quad T_u \ge \frac{\kappa(A^2)}{\|A^2\|} \log \frac{1}{\delta} = \Big(\frac{\kappa}{\|A\|} \Big)^2 \log \frac{1}{\delta} , \]
where the truncated evolution time depends poorly on the condition number $\kappa$ of $A$. In contrast, as shown below, the LCHS in \eqref{abstractinvA} for the QLSP achieves a linear dependence on the condition number.

\begin{theorem} \label{thm:underlyingODE}
Let $A$ be an invertible Hermitian matrix and define
\begin{equation}\label{xT}
\bb{x}_T = \int_0^T \int_{\mathbb{R}}  \varphi(k) \e^{-\i k A s} \d k \d s \bb{b}.
\end{equation}
If $T = \Theta(\frac{\kappa}{\|A\|}  \sqrt{2 \log\frac{1}{\delta}})$, where $\kappa$ is the condition number of $A$ and $\delta>0$ is a constant, then there holds
\[ \|\bb{x} - \bb{x}_T\| \le \delta\|\bb{x}\|.\]
Further, we have $\bb{x}_T = \bb{u}(T')$, where $T' = T^2/2$ and $\bb{u}$ satisfies \eqref{underlyingODE}.

\end{theorem}
\begin{proof}
The diagonalization of $A$ yields
\begin{align*}
\|\bb{x} - \bb{x}_T\|
		&= \Big\|\int_0^{\infty} \int_{\mathbb{R}}  \zeta(k) \e^{-\i k A s} \d k \d s\bb{b}
		- \int_0^T \int_{\mathbb{R}}  \zeta(k) \e^{-\i k A s} \d k \d s\bb{b}\Big\| \\
		&\le \max_j \Big|\lambda_j \Big(\int_0^\infty \hat{\varphi}(\lambda_js) \d s - \int_0^T \hat{\varphi}(\lambda_js) \d s\Big) \Big| \|\bb{x}\| \\
		&= \max_j \e^{-(\lambda_jT)^2/2}  \|\bb{x}\| \le \delta \|\bb{x}\|,
	\end{align*}
if we set $T = \Theta(\frac{\kappa}{\|A\|}  \sqrt{2 \log\frac{1}{\delta}})$, where we have used the fact that
\[\int_0^t \hat{\varphi}(\lambda_js) \d s = \frac{1}{\lambda_j}(1 - \e^{-(\lambda_jt)^2/2}). \]
This also implies
\begin{align*}
 \int_0^T \int_{\mathbb{R}}  \varphi(k) \e^{-\i k A s} \d k \d s \bb{b}
 = (I - \e^{-(AT)^2/2}) A^{-1} \bb{b}
  = \int_0^{T^2/2} \e^{-A^2 (T-s) }\d s \bb{b} ,
\end{align*}
as required.
\end{proof}

\subsection{Schr\"odingerization-form problems for matrix inverse}\label{sec:SchrodingerizationQLSP}
	
In the following, we show that the LCHS formula in Theorem  \ref{thm:abstractLCHSQLSA} can be converted into the LCHS with the Hamiltonian given by convection equations in one higher dimension. This transformation allows us to apply the discrete Fourier transform, similar to the approach used in the Schr\"odingerization method for ODEs.

\begin{theorem}\label{thm:discreteFourier}
Let $ A $ be an invertible Hermitian matrix. Then there holds
\begin{equation}\label{integralt}
	\bb{x} = A^{-1}\bb{b} = \int_0^\infty \bb{v}(t,0) \d t,
\end{equation}
where $\bb{v}(t,p)$ satisfies the following system of convection equations
\begin{equation}\label{vtp}
\begin{cases}
\dfrac{\d }{\d t} \bb{v}(t,p) = A  \partial_p \bb{v} (t,p), \\
\bb{v}(0,p) =  \zeta(p) \bb{b}
\end{cases}
\end{equation}
with $\zeta(p) = \hat{\varphi}(p) = p \e^{-p^2/2}$.
\end{theorem}

\begin{proof}
According to Theorem \ref{thm:abstractLCHSQLSA}, we have
\[A^{-1} \bb{b} =  \int_0^{\infty} \int_{\mathbb{R}}  \e^{-\i k A s} (\varphi(k)\bb{b} ) \d k \d s. \]
Assume that $\bb{v}(t,p)$ is the Fourier transform of $\check{\bb{v}}(t,k) = \e^{-\i k A t} (\varphi(k)\bb{b} )$,
namely,
\[\bb{v}(t,p) = \int_{\mathbb{R}} \e^{-\i k p} \check{\bb{v}}(t,k) \d k =  \int_{\mathbb{R}} \e^{-\i k p} \e^{-\i k A t} (\varphi(k)\bb{b} ) \d k.\]
Noting that $\check{\bb{v}}(t,k)$ satisfies
\[\begin{cases}
\dfrac{\d }{\d t} \check{\bb{v}}(t,k) = -\i k A \check{\bb{v}}(t,k), \qquad \check{\bb{v}}(0,k) = \varphi(k)\bb{b},\\
	\lim_{|k|\to +\infty} \check{\bb{v}}(t,k) = \bb{0},
\end{cases}\]
we apply the Fourier transform to get
\[\begin{cases}
	\partial_t \bb{v}(t,p) = A  \partial_p \bb{v} (t,p), \\
	\bb{v}(0,p) = \int_{\mathbb{R}} \e^{-\i k p}\varphi(k) \d k \bb{b} = \hat{\varphi}(p) \bb{b} = p \e^{-p^2/2} \bb{b}.
\end{cases}\]
It is clear that the solution $\bb{x}$ can be restored by
\[\bb{x} = A^{-1}\bb{b} = \int_0^\infty \bb{v}(t,0) \d t.\]
This completes the proof.
\end{proof}

The Schr\"odingerization method in \cite{JLY22SchrShort, JLY22SchrLong} transforms linear non-unitary dynamics into Schr\"odinger-type systems via the so-called warped phase transformation, which maps the equation into one higher dimension and results in a system of linear convection equations such as \eqref{vtp}. For this reason, we refer to \eqref{vtp} as a Schr\"odingerization-form problem. Theorem \ref{thm:discreteFourier} then shows that the solution $\bb{x}$ can be expressed as the LCHS of such problems.

Since $\zeta(p)$ is odd and decays rapidly as $|p|\to\infty$, we can truncate the domain to a finite interval $p\in[-R_0,R_0]$ with $R_0$ sufficiently large, so that the truncated problem can be treated with periodic boundary conditions up to a negligible error. To realize the exact periodic boundaries, we may employ the cut-off function technique described below.

We begin by recalling the mollifier, defined as
\begin{equation}\label{mollifier}
\eta(p) = \begin{cases}
  \frac{1}{C}\exp \left( \frac{1}{| p |^2 - 1} \right), & \quad | p | < 1  \\
  0, & \quad | p | \ge 1
\end{cases}, \qquad C = \int_{B_1} \exp\left(\frac{1}{|p|^2 - 1}\right) \mathrm{d}p,
\end{equation}
where $ B_1 $ denotes the unit ball in $\mathbb{R}^n$ and $ C $ is the normalization constant ensuring $ \int_{\mathbb{R}^n} \eta(p)  \d p = 1 $. This function belongs to $ C_0^\infty(\mathbb{R}^n) $ with support $ \overline{B_1} $.

\begin{lemma}\label{lem:etak}
The mollifier satisfies the following estimate for its derivatives in one dimension:
\begin{equation}\label{mollifierbound}
|\eta^{(k)} (p) | \lesssim C(k): = 20^k k!\e^{-2k} (2k)^{2k},\quad \forall p\in \mathbb{R}.
\end{equation}
\end{lemma}

For any $ \varepsilon > 0 $, we can rescale the function such that its support becomes $ \overline{B_\varepsilon} $, a closed ball of radius $ \varepsilon $. The rescaled function is given by $\eta_\varepsilon(p) = \frac{1}{\varepsilon^n} \eta \left( \frac{p}{\varepsilon} \right)$.
For a function $ u \in L_{\text{loc}}^1(\Omega) $, the mollifier operator $ J_\varepsilon $ is defined through convolution as
\begin{equation}\label{eq:mollifier1}
 J_\varepsilon u(p) = (\eta_\varepsilon * u)(p) = \int_{\Omega} \eta_\varepsilon(p - y) u(y) \d y
= \int_{B_\varepsilon(p)} \eta_\varepsilon(p - y) u(y) \d y, \quad p \in \Omega_\varepsilon,
\end{equation}
where the domain $ \Omega_\varepsilon $ is defined by
\[ \Omega_\varepsilon = \left\{ p \in \Omega : \overline{B_\varepsilon(p)} \subset \Omega \right\} = \left\{ p \in \Omega : \text{dist}(p,\partial \Omega) > \varepsilon \right\}. \]
It can be verified that $ J_\varepsilon u \in C^{\infty}(\Omega_\varepsilon) $ for every $ u \in L_{\text{loc}}^1(\Omega) $. Furthermore, if $ \text{supp} \{ u \} \Subset \Omega $, denoting $ \delta = \text{dist}(\text{supp} \{ u \}, \partial \Omega) $, then for $ \varepsilon < \delta/4 $, we have $ J_\varepsilon u \in C_0^\infty (\Omega) $ with $ \text{supp} \{ J_\varepsilon u \} \subset \Omega_\varepsilon $.

Now we are ready to describe the construction of the cut-off function.

\begin{lemma}[cut-off function] \label{lem:cutoff}
Let $\Omega \subset \mathbb{R}^n$ be a non-empty open set, and ${\Omega _0} \Subset \Omega $. Define
\[
\delta = \text{dist}(\Omega _0, \partial \Omega), \quad d = \frac{\delta}{4}, \quad \Omega_1 = \{ p \in \Omega : \text{dist}(p, \Omega _0) < d \}.
\]
Let $\chi _{\Omega _1}(p)$ denote the indicator function of $\Omega_1$. Then $\rho = J_d \chi _{\Omega _1} $ satisfies
\[
\begin{cases}
 \rho \in C_0^\infty (\Omega ), & \text{supp} \{ \zeta \} \subset K_d, \\
\rho (p) \equiv 1, & p \in \Omega _0, \\
0 \le \rho (p) \le 1, & p \in \Omega ,
\end{cases}
\]
where
\[
K_d = \{ p \in \Omega : \text{dist}(p, \Omega _1) \le d \} = \{ p \in \Omega : \text{dist}(p, \Omega _0) \le 2d \}.
\]
The function $\rho$ is referred to as the cut-off function relative to the subset $\Omega _0$ in $\Omega$.
\end{lemma}

The one-dimensional cut-off function satisfies the following estimate for its derivatives:
\begin{equation}\label{zetak}
| \rho^{(k)}(p) | \lesssim \frac{C(k)}{d^k},
\end{equation}
where $C(k)$ is defined in \eqref{mollifierbound}.

For our problem, we set $\Omega_0 = (-R_0,R_0)$ and $d \ge 1$. Let
\begin{equation}\label{cutoffmethod}
    \psi(p) = \rho(p)\zeta(p),
\end{equation}
where $\rho(p)$ is the cut-off function defined in Lemma \ref{lem:cutoff}. Then, $\psi(p) = \zeta(p)$ on $[-R_0, R_0]$ and $\text{supp}\{\psi\} \subset [-R_1, R_1]$ with $R_1 = R_0+2d$. In place of \eqref{vtp}, we now consider a modified Schr\"odingerization-form problem
\begin{equation} \label{modifiedSchrProb}
\begin{cases}
\partial_t \bb{v}^{\text{cut}}(t,p) = A  \partial_p \bb{v}^{\text{cut}} (t,p),\\
\bb{v}^{\text{cut}}(0,p) =  \psi(p) \bb{b}.
\end{cases}
\end{equation}
It is clear that $\bb{v}^{\text{cut}}(0,-R_1) =  \bb{v}^{\text{cut}}(0,R_1) = \bb{0}$, however, this does not imply $\bb{v}^{\text{cut}}(t,-R_1) =  \bb{v}^{\text{cut}}(t,R_1) = \bb{0}$ for all $t\in [0,T]$. In view of the transport property, we can truncate \eqref{modifiedSchrProb} to a larger domain $[-R, R]$ with
\begin{equation}\label{Rnew}
R = R_1 + \lambda_{\text{smax}}T = R_0 + 2d + \lambda_{\text{smax}}T,
\end{equation}
where $\lambda_{\text{smax}} \ge \|A\|$ is greater than the largest absolute value among the eigenvalues of $A$. Moreover, if $R_0 > \lambda_{\text{smax}}T$, then by the method of characteristics one can verify that
\[\bb{v}^{\text{cut}}(t,p) = \bb{v}(t,p), \qquad \forall p \in [ - R_f, R_f], \quad R_f = R_0 - \lambda_{\text{smax}}T.\]

\begin{lemma}\label{lem:truncSchr}
Let $R$ be as defined in \eqref{Rnew}. Suppose that $\bb{w}(t,p)$ is the solution of the following periodic truncation problem
\begin{equation} \label{perExtension}
\begin{cases}
\partial_t \bb{w}(t,p) = A  \partial_p \bb{w} (t,p), \quad t\in (0,T), ~~ p \in (-R, R),\\
\bb{w}(0,p) =  \psi^{\text{per}}(p)\bb{b},  \quad p \in [-R, R]\\
\bb{w}(t,-R) = \bb{w}(t,R),  \quad t \in [0,T],
\end{cases}
\end{equation}
where $\psi^{\text{per}}$ is the periodic extension of $\psi(p)$ from $[-R, R]$.
\begin{enumerate}
  \item[(1)] If $R_f = R_0 - \lambda_{\text{smax}}T>0$, then
\[\bb{w}^{(k)}(t,p)  = \bb{0}, \quad  p = \pm R,\]
\[\bb{w}^{(k)}(t,p)  = \bb{v}^{(k)}(t,p), \qquad \forall p \in [-R_f,R_f]. \]
where $\bb{v}$ is the solution of \eqref{vtp}.
  \item[(2)] If $d = r\ge 1$, then
  \begin{equation}\label{psibound}
  \|(\psi^{\text{per}})^{(r)}\|_{L^2((-R,R))}^{1/r} \le \|\psi^{(r)}\|_{L^2(\mathbb{R})}^{1/r} \lesssim r^3.
  \end{equation}
\end{enumerate}
\end{lemma}

\begin{proof}
The result in (1) is a direct consequence of the method of characteristics.
For the second result, by setting $\xi_k(p) = \rho^{(r-k)}(p) \zeta^{(k)}(p)$, we have
\begin{align*}
|\psi^{(r)}(p)|
& = \Big|\sum_{k=0}^r C_r^k \rho^{(r-k)}(p) \zeta^{(k)}(p)\Big| = \Big|\sum_{k=0}^r C_r^k \xi_k(p)\Big|\\
& \le ((C_r^0)^2 + \cdots +(C_r^r)^2)^{1/2} ( |\xi_0|^2 + \cdots + |\xi_r|^2 )^{1/2}  \\
& = (C_{2r}^r)^{1/2} ( |\xi_0|^2 + \cdots + |\xi_r|^2 )^{1/2} \le 2^r ( |\xi_0|^2 + \cdots + |\xi_r|^2 )^{1/2}
\end{align*}
for any $p\in\mathbb{R}$.

Let $\text{erf}(p)$ be the error function, which is defined as
\[\text{erf}(p) = \frac{2}{\sqrt{\pi}} \int_0^p \e^{-t^2} \d t.\]
Since $\text{erf}''(p) = -\frac{4}{\sqrt{\pi}} p\e^{-p^2}$, we have
\[\zeta(p) = - \frac{1}{4}\sqrt{\frac{\pi}{2}}\text{erf}''\Big( \frac{p}{\sqrt{2}} \Big).\]
Noting that
\[\text{erf}^{(k)}(p) = \frac{2}{\sqrt{\pi}} (-1)^{k-1} H_{k-1}(p) \e^{-p^2}, \qquad k \ge 1,\]
where $H_k$ is the Hermitian polynomial,  defined by $H_k(p) = (-1)^k \e^{p^2} (\e^{-p^2})^{(k)}$, which leads to
\[\zeta^{(k)}(p) = - \frac{1}{4}\sqrt{\frac{\pi}{2}} \frac{1}{(\sqrt{2})^k}\text{erf}^{(k+2)}\Big( \frac{p}{\sqrt{2}} \Big).\]
It is known that
\[
|H_k(p)| \le C \, 2^{k/2} \sqrt{k!} \e^{p^2/2},
\]
where $C \approx 1.086435$ (see \cite[Eq.~(7.66)]{Shenspectral} for example). We then obtain
\begin{align*}
|\zeta^{(k)}(p)|
\le \frac{1}{2} \cdot \frac{1}{(\sqrt{2})^{k+1}} \, \bigl| H_{k+1}(p/\sqrt{2}) \bigr| \e^{-p^2/2}
\le \frac{C}{2} \sqrt{(k+1)!} \e^{-p^2/4}.
\end{align*}
By setting $d=r$, this along with \eqref{zetak} yields
\begin{align*}
|\xi_k(p)|
& \lesssim \frac{20^{r-k} (r-k)!\e^{-2(r-k)} (2(r-k))^{2(r-k)}}{d^{r-k}} \sqrt{(k+1)!} \e^{-p^2/4} \\
& \le \frac{80^{r-k} (r-k)^{3(r-k)}\e^{-2(r-k)}}{d^{r-k}} \sqrt{(k+1)!} \e^{-p^2/4} \\
& = \Big( \frac{80 \e^{-2} (r-k)^3}{r} \Big)^{r-k}  \sqrt{(k+1)!} \e^{-p^2/4}
\lesssim (80 \e^{-2})^r r^{5 r/2}  \e^{-p^2/4},
\end{align*}
leading to
\[|\psi^{(r)}(p)| \lesssim 2^r r^{1/2} \max_{0\le k \le r} |\xi_k| \le (80 \e^{-2})^r 2^r r^{3r} \e^{-p^2/4}. \]
Therefore, we obtain
\begin{align*}
\|\psi^{(r)}\|_{L^2(\mathbb{R})}^{1/r} = \Big(\int_{\mathbb{R}} |\psi^{(r)}(p)|^2 \d p \Big)^{1/(2r)}
\lesssim r^3.
\end{align*}
This completes the proof.
\end{proof}

Based on the above result, we can develop a quantum algorithm utilizing the discrete Fourier transform. Unlike the direct solution for an ODE problem, we need to address the integral with respect to the ``artificial time'' in \eqref{integralt}.

\section{Implementation} \label{sec:implementationSchrQLSP}

We focus solely on Hermitian coefficient matrix. For a non-Hermitian matrix $A$, we introduce the dilation matrix $ \tilde{A} = \ket{0}\bra{1}\otimes A + \ket{1}\bra{1}\otimes A^\dag$, which is Hermitian.

\subsection{Discretization of the auxiliary variable}	

Let $p\in [a,b]= [-R, R]$ with $R>0$ satisfying \eqref{Rnew} and consider the periodic truncation problem \eqref{perExtension}. Then one can apply the Fourier spectral method by discretizing the $p$ domain.  Toward this end,  we choose a uniform mesh size $\Delta p = (b-a)/N_p$ for the auxiliary variable with $N_p=2^{n_p}$ being an even number. The grid points are denoted by $a = p_0<p_1<\cdots<p_{N_p} = b$.
For $p\in [a, b]$, the one-dimensional basis functions are usually chosen as
\[\phi_l(p) = \e ^{\i \mu_l (p-a)} , \quad \mu_l = \frac{2\pi (l-N_p/2) }{b-a}, \quad l=0,1,\cdots,N_p-1.\]
For later uses, we let
\[\mu_{\max} = \max_l |\mu_l| = \frac{\pi N_p}{b-a} = \frac{\pi N_p}{2R} = \frac{\pi}{\Delta p}.\]
We define the complex discrete Fourier space by
\[S_p = \text{span} \{\phi_l(p): l = 0,1,\cdots, N_p-1 \}\]
and assume $u \in C_p[a,b]$, which is a continuous and periodic function on $[a,b]$.

The Fourier interpolation $u_I\in S_p$ is defined by $u_I(p_k) = u(p_k)$ for $k = 0,1,\cdots, N_p-1$. Let
\begin{equation}\label{uInterpolation}
	u_I(p) = \sum\limits_{l= 0}^{N_p-1} \tilde{u}_l \phi_l(p).
\end{equation}
The coefficients can be explicitly written as
\[\tilde{u}_l = \frac{1}{N_p} \sum\limits_{k=0}^{N_p-1} u(p_k) \e ^{ - \i \mu_l (p_k-a)} .\]
The interpolation can be written in matrix form as $\bb{u}_I = \bb{u} = \Phi \tilde{\bb{u}}_l$, where
\[\bb{u}(t) = (u(p_j))_{N_p\times 1}, \quad \tilde{\bb{u}}_l = (\tilde{u}_l)_{N_p\times 1}, \quad
\Phi = (\phi_{jl})_{N_p\times N_p} = (\phi_l(p_j))_{N_p\times N_p}.\]
The momentum operator $\hat{p} = -\i \partial_p $ can be discretized as
\begin{align*}
	\hat{p}u(p)
	 \approx  \sum\limits_{l=0}^{N_p-1} \tilde{u}_l (-\i \partial_p \phi_l(p))
	 = \sum\limits_{l=0}^{N_p-1} \tilde{u}_l \mu_l \phi_l(p)
\end{align*}
for $p = p_j$, $j = 0,1,\cdots, N_p-1$, which is written in matrix form as
\[\hat{p}^{\rm d} \bb{u} =  \Phi D_\mu \Phi^{-1} \bb{u} =: P_\mu \bb{u}, \qquad
D_\mu = \text{diag} ( \mu_0, \cdots, \mu_{N_p-1} ),\]
where $\hat{p}^{\rm d}$ is the discrete momentum operator.

We denote by $\Pi u$ the $L^2$ projection onto $S_p$, namely,
\[(\Pi u , v) = (u , v), \quad v \in S_p.\]
One can verify that
\[
\Pi u (p) = \sum\limits_{l= 0}^{N_p-1} \hat{u}_l \phi_l(p), \quad \hat{u}_l = \frac{1}{b-a} \int_a^b u(p) \e ^{-\i \mu_l (p-a)} \d p, \quad l = 0, \cdots, N_p-1.
\]
For the projection, we have the following approximation error in the maximum norm.

\begin{lemma}\label{lem:L2errspectral}
	For any $m>1/2$ and $u \in C_p^m[a,b]$, there exists a positive constant $C$, independent of $N_p$, such that
	\[\|u - \Pi u\|_{L^\infty} \le C \Big(\frac{b-a}{N_p}\Big)^{m-1/2} \|u^{(m)}\|_{L^2[a,b]}.\]
	Here, $C_p^m[a,b]$ consists of functions with derivatives of order up to $(m-1)$ being periodic on $[a,b]$.
\end{lemma}
The proof was given in \cite[Theorem 2.12]{HGG2007} for $[a,b] = [0,2\pi]$. Scaling argument yields the desired result. For $m>1$, the space $H^m( (a,b) )$ can be embedded in $L^\infty((a,b))$. Hence, the conclusion also holds for $u \in H_p^m[a,b]$ with $m>1$.

Let $\bb{w}(t,p) = [w_1(t,p), \cdots, w_N(t,p)]^\top$ be the solution to \eqref{perExtension}. The spectral discretization of $w_i(t,p)$ is
\[w_{i,h}(t,p) = \sum_{l=0}^{N_p-1} \tilde{w}_{i,l,h}(t) \phi_l(p), \qquad i = 1,\cdots,N,\]
where we use the subscript $h$ to denote the numerical solution for \eqref{vtp}. The approximate solution is then given by $\bb{w}_h(t,p) = [w_{1,h}(t,p),\cdots, w_{N,h}(t,p)]^\top$, which can be written as
\begin{equation}\label{interpcoeff}
\bb{w}_h(t,p) = \sum_{l=0}^{N_p-1} \tilde{\bb{w}}_{l,h}(t) \phi_l(p), \qquad
\tilde{\bb{w}}_{l,h}(t) = \frac{1}{N_p} \sum\limits_{k=0}^{N_p-1} \bb{w}_h(t,p_k) \e ^{ - \i \mu_l (p_k-a)} .
\end{equation}
Let the vector $\bb{W}_h$ be the collection of the function $\bb{w}_h$ at the grid points, defined more precisely as
\[\bb{W}_h(t) = \sum_{k,i} w_{i,h}(t,p_k) \ket{k,i}
= [\bb{w}_h(t,p_0) ; \cdots; \bb{w}_h(t,p_{N_p-1})].\]
Accordingly, we define
\[\tilde{\bb{W}}_h(t) = \sum_{l,i} \tilde{w}_{i,l,h}(t) \ket{l,i}
= [\tilde{\bb{w}}_{0,h}(t) ; \cdots; \tilde{\bb{w}}_{N_p-1,h}(t)],\]
with ``;'' indicating the straightening of $\{\tilde{\bb{w}}_{i,h}\}_{i\ge 1}$ into a column vector.
One can find that $\bb{W}_h(t) = (\Phi \otimes I_{N\times N}) \tilde{\bb{w}}_h(t)$.
The equation in \eqref{vtp} is then transformed into
\begin{equation}\label{wh}
	\begin{cases}
		\dfrac{\d }{\d t} \bb{W}_h(t) = \i ( P_\mu \otimes A ) \bb{W}_h(t), \\
		\bb{W}_h(0) = \bb{\psi} \otimes \bb{b},
	\end{cases}
\end{equation}
where $\bb{\psi} = [\psi(p_0), \cdots, \psi(p_{N_p-1})]^\top = \bb{\psi}^{\text{per}}$.
In terms of $\tilde{\bb{W}}_h = (\Phi^{-1} \otimes I)\bb{W}_h$, one gets the following Hamiltonian system:
\begin{equation}\label{discreteLCHSQLSP}
\begin{cases}
\dfrac{\d}{\d t} \tilde{\bb{W}}_h(t) = \i (D_\mu \otimes A) \tilde{\bb{W}}_h(t), \\
\tilde{\bb{W}}_h(0) = \tilde{\bb{\psi}} \otimes \bb{b}, \quad \tilde{\bb{\psi}} = \Phi^{-1}\bb{\psi}.
\end{cases}
\end{equation}

\subsection{Truncation of the integral}

The solution to \eqref{discreteLCHSQLSP} can be written as
\begin{equation}\label{Axb}
	\tilde{\bb{W}}_h(t) = \e^{\i (D_\mu \otimes A)t} \tilde{\bb{W}}_h(0),
\end{equation}
which gives
\[\bb{W}_h(t)  = [\bb{w}_h(t,p_0) ; \cdots; \bb{w}_h(t,p_{N_p-1})] = (\Phi \otimes I)\e^{\i (D_\mu \otimes A)t} \tilde{\bb{W}}_h(0). \]
According to Theorem \ref{thm:discreteFourier}, we can recover the approximate solution to the linear systems problem by projecting $\bb{W}_h(t)$ onto $\ket{p=0}$ and truncate the integral as
\begin{equation}\label{xhT}
	\bb{x}_{h,T} = \int_0^T \Pi_*\bb{W}_h(t) \d t = \int_0^T \bb{w}_h(t,0) \d t,
\end{equation}
where $\Pi_* = \ket{k_*}\bra{k_*} \otimes I$ with $k_*$ satisfying $p_{k_*} = 0$.

\begin{theorem}\label{thm:pdiscretization}
Let $p\in [-R, R]$ with $R>0$ satisfying \eqref{Rnew}. Then for any $r>1$, there holds
\begin{equation}\label{errorpk}
|\bb{w}(\cdot,p_k) - \bb{w}_h(\cdot,p_k)| \lesssim (\Delta p)^{r-1/2} \|\psi^{(r)}\|_{L^2((-R,R))} \|\bb{b}\|, \quad k = 0,1,\cdots,N_p-1.
\end{equation}
For any given $0<\delta<1$, let $T = \Theta(\frac{\kappa}{\|A\|}  \sqrt{2 \log\frac{1}{\delta}})$. If $r\ge 2$ and $r \simeq \log(1/\delta)$ and the mesh size $\triangle p$ satisfies
\[(\Delta p)^{-1} \simeq \mu_{\max} \simeq \log^3 \frac{\kappa }{\xi \|A\| \delta},\]
then there holds
\begin{equation}\label{error2}
\|\bb{x} - \bb{x}_{h,T}\| \lesssim \delta \|\bb{x}\|.
\end{equation}
\end{theorem}

\begin{proof}
	(1) We first prove \eqref{errorpk}.
	For the exact solution $\bb{w}$ to \eqref{vtp}, the $L^2$ projection $\Pi \bb{w}$ is defined as $\Pi \bb{w}= [\Pi w_1, \cdots, \Pi w_n]^\top$. We can write it as
	\begin{equation}\label{projectioncoeff}
		\Pi \bb{w}(t,p) = \sum\limits_{l= 0}^{N_p-1} \hat{\bb{w}}_l(t) \phi_l(p), ~~ \hat{\bb{w}}_l(t) = \frac{1}{b-a} \int_a^b \bb{w}(t,p) \e ^{-\i \mu_l (p-a)} \d p, ~~ l = 0, \cdots, N_p-1,
	\end{equation}
where $a = -R$ and $b = R$. By the triangle inequality, the error can be split as
	\[|\bb{w}(\cdot,p_k) - \bb{w}_h(\cdot,p_k)|
	\le |\bb{w}(\cdot,p_k) - \Pi \bb{w}(\cdot,p_k)| + |\Pi \bb{w}(\cdot,p_k) - \bb{w}_h(\cdot,p_k)|.\]
When $R$ satisfies \eqref{Rnew}, $\bb{w}^{(r)}(t,p) = \bb{0}$ at $p = a, b$. This implies that each entry of $\bb{w}(t,p)$ can be treated as a function in $C_p^r[a,b]$ for any $r\ge 0$.
	
For the first term on the right-hand side, by Lemma \ref{lem:L2errspectral}, there holds
	\begin{align*}
		|\bb{w}(\cdot,p_k) - \Pi \bb{w}(\cdot,p_k)|
		& \le \|\bb{w}(\cdot,p) - \Pi \bb{w}(\cdot,p)\|_{L^{\infty}((a,b))}
		\lesssim (\Delta p)^{r-1/2} \|\bb{w}^{(r)}(\cdot,p)\|_{L^2((a,b))} \\
		& \lesssim (\Delta p)^{r-1/2} \|\psi^{(r)}\|_{L^2((a,b))} \|\bb{b}\|.
	\end{align*}
For the second term, by definition, one gets
	\[\Pi \bb{w}(\cdot,p) - \bb{w}_h(\cdot,p) = \sum\limits_{l= 0}^{N_p-1} (\hat{\bb{w}}_l(t)- \tilde{\bb{w}}_{l,h}(t)) \phi_l(p). \]
	The evolution of $\tilde{\bb{w}}_{l,h}(t)$ is given in \eqref{discreteLCHSQLSP}. For each $\hat{\bb{w}}_l(t)$, we obtain from its definition and \eqref{vtp} that
	\[\dfrac{\d }{\d t} \hat{\bb{w}}_l(t) = \i \mu_l A  \hat{\bb{w}}_l(t), \quad l = 0,\cdots, N_p-1.\]
	Let $\hat{\bb{W}}(t) = \sum_{l,i} \hat{v}_{i,l}(t) \ket{l,i} = [\hat{\bb{w}}_0(t) ; \cdots; \hat{\bb{w}}_{N_p-1}(t)]$. The above system can be rewritten as
	\[\dfrac{\d }{\d t} \hat{\bb{W}}(t) = \i (D_\mu \otimes A)  \hat{\bb{W}}(t).\]
	Introducing the vector
	\[\bb{W}_{\pi}(t) = (\Phi \otimes I) \hat{\bb{W}}(t) = [\Pi \bb{w}(t,p_0); \cdots; \Pi \bb{w}(t,p_{N_p-1})],\]
	we have
	\[\begin{cases}
		\dfrac{\d }{\d t} \bb{W}_{\pi}(t) = \i ( P_\mu \otimes A ) \bb{W}_{\pi}(t), \\
		\bb{W}_{\pi}(0) = \bb{\psi}_\pi \otimes \bb{b},
	\end{cases}\]
	where $\bb{\psi}_\pi = [\Pi \psi(p_0), \cdots, \Pi \psi(p_{N_p-1})]^\top$. Let $\bb{e}_h(t) =  \bb{W}_h(t) - \bb{W}_\pi (t)$. Then,
	\[\begin{cases}
		\dfrac{\d }{\d t} \bb{e}_h(t) = \i ( P_\mu \otimes A ) \bb{e}_h(t), \\
		\bb{e}_h(0) = (\bb{\psi} - \bb{\psi}_\pi) \otimes \bb{b},
	\end{cases}\]
	which gives $\bb{e}_h(t) = \e^{\i ( P_\mu \otimes A ) t}  \bb{e}_h(0)$ and
	\[|\bb{w}_h(\cdot,p_k) - \Pi \bb{w}(\cdot, p_k)| \le \|\bb{e}_h(t)\| = \|\bb{\zeta} - \bb{\zeta}_\pi\| \cdot \|\bb{b}\|
	\lesssim (\Delta p)^{r-1/2} \|\psi^{(r)}|_{L^2((a,b))} \|\bb{b}\|.\]
	
(2) Let $\bb{x}_T$ be defined in \eqref{xT}. According to Theorem \ref{thm:underlyingODE}, one has
\[\|\bb{x} - \bb{x}_{h,T}\| \le \|\bb{x} - \bb{x}_T\| + \|\bb{x}_T - \bb{x}_{h,T}\| \le \delta\|\bb{x}\| + \|\bb{x}_T - \bb{x}_{h,T}\|,\]
provided that $T = \Theta(\frac{\kappa}{\|A\|}  \sqrt{2 \log\frac{1}{\delta}})$. By definition,
\begin{align*}
\bb{x}_T - \bb{x}_{h,T}
 = \int_0^T \bb{v}(t,0) \d t - \int_0^T \bb{w}_h(t,0) \d t
 = \int_0^T \bb{w}(t,0) \d t - \int_0^T \bb{w}_h(t,0) \d t.
\end{align*}
where $\bb{v}$ and $\bb{w}$ are the solutions to \eqref{vtp} and \eqref{perExtension}, respectively.
Using the estimation established in the first step and Eq.~\eqref{psibound}, we obtain
\begin{align*}
\|\bb{x}_T - \bb{x}_{h,T}\|
& \lesssim (\Delta p)^{r-1/2} \|\psi^{(r)}\|_{L^2((a,b))}  T \|\bb{b}\|
 \lesssim (\Delta p)^{r-1} (r^3)^r  T \|\bb{b}\| \\
& \lesssim (\Delta p)^{r-1} (r^3)^r  \frac{\kappa}{\|A\|}  \sqrt{\log\frac{1}{\delta}}  \|\bb{b}\|
 = (\Delta p)^{r-1} (r^3)^r  \frac{\kappa \|A^{-1}\bb{b}\|}{\|A\| \|A^{-1}\ket{b}\|}  \sqrt{\log\frac{1}{\delta}} \\
& = (\Delta p)^{r-1} (r^3)^r  \frac{\kappa }{\xi \|A\|}  \sqrt{\log\frac{1}{\delta}} \|\bb{x}\|,
\end{align*}
where $\xi = \|A^{-1}\ket{b}\|$ and $\Delta p<1$. Thus, we can require
\[(\Delta p)^{r-1} (r^3)^r  \frac{\kappa }{\xi \|A\|}  \sqrt{\log\frac{1}{\delta}} \simeq \delta \quad \mbox{or}
\quad \Delta p   \simeq \frac{1}{r^{3r/(r-1)}}\Big( \frac{\delta}{\sqrt{\log\frac{1}{\delta}} \frac{\kappa }{\xi \|A\|}} \Big)^{1/(r-1)}.\]
For sufficiently large $r$, one has
\[\e \le \Big( \frac{\delta}{\sqrt{\log\frac{1}{\delta}} \frac{\kappa }{\xi \|A\|}} \Big)^{1/(r-1)} \le \e^2, \qquad
\frac12 \le \frac{1}{r-1} \Big( \log \frac{\kappa }{\xi \|A\| \delta} + \frac12 \log \log \frac{1}{\delta} \Big) \le 1.\]
Noting that
\[\log \frac{\kappa }{\xi \|A\| \delta} \simeq \Big( \log \frac{\kappa }{\xi \|A\| \delta} + \frac12 \log \log \frac{1}{\delta} \Big),\]
\[r \le 2 (r-1), \qquad r\ge 2,\]
where we have used $\frac{\kappa }{\xi \|A\|} \ge 1$, we then obtain
\[r \simeq (r-1) \simeq \log \frac{\kappa }{\xi \|A\| \delta} \quad \mbox{and} \quad
\Delta p \simeq \frac{1}{\log^3 \frac{\kappa }{\xi \|A\| \delta}} . \]
This completes the proof.
\end{proof}

\subsection{Discretization of the truncated integral}

Let $U(A,t)=\e^{\i (D_\mu \otimes A)t}$. We have
\begin{equation}\label{xTtruncate}
\bb{x} \approx \bb{x}_T \approx \bb{x}_{h,T} = \Pi_* (\Phi \otimes I) g(A)\tilde{\bb{W}}_h(0), \qquad g(A) = \int_0^T U(A,t) \d t
\end{equation}
for sufficiently large $T$, where $\Pi_*$ is the projector defined in \eqref{xT}.

To present an explicit algorithm, we need to express the truncated integral as a finite sum using numerical integration.
Let $t_m = m \tau$ for $m=0, 1,\cdots,N_t$, where $N_t = T/\tau$ and $\tau$ is the step size. We use composite Gaussian quadrature to discretize the variable $t$ and obtain
\begin{equation}\label{ShrodingerhA}
	g(A) = \sum_{m=0}^{N_t-1} \int_{m\tau}^{(m+1)\tau} U(A,t) \d t
	\approx \sum_{m=0}^{N_t-1} \sum_{q=0}^{Q-1} c_{m,q} \e^{\i (D_\mu \otimes A)t_{m,q}} =: p(A),
\end{equation}
where, on each interval $[m \tau, (m + 1)\tau]$, we use Gaussian quadrature with $Q$ nodes. The $t_{m,q}$'s are the Gaussian nodes, $c_{m,q} = w_{q}$ and $w_{q}$'s are the Gaussian weights (which do not depend on the choice of $m$).

In the implementation, we assume that the Hamiltonian simulation $\e^{\i (D_\mu \otimes A)t}$ has error at most $\delta_2$ for $t\in [0,T]$, and we denoted its approximation by $V_{m,q}$. The resulting approximation of $p(A)$ is denoted by $\tilde{p}(A)$. Then, the approximation for \eqref{xTtruncate} is
\begin{equation}\label{xTtruncateGauss}
\bb{x}_{h,T}^{\text{d}} =: \Pi_* (\Phi \otimes I) \tilde{p}(A)\tilde{\bb{W}}_h(0).
\end{equation}

To measure the accuracy of the approximation, we first introduce the quadrature error described as follows.
\begin{lemma}\label{lem:ShrodingerGaussErr}
	Let $A$ be a Hermitian matrix. Suppose that $f(\cdot,x)$ is a function in $C^{2Q}[a,b]$ with respect to $x$ and $f(A,x)$ is well-defined. Let $w_q$ and $x_q$ be the Gaussian quadrature weights and points on $[a,b]$ for $q=0,1,\cdots,Q-1$. There there holds
	\[\Big\|\int_a^b f(A,x) \d x - \sum_{q=0}^{Q-1} w_q f(A,x_q) \Big\| \le  \frac{(b-a)^{2Q+1}(Q!)^4 }{(2Q+1)[ (2Q)! ]^3} \max_{x\in [a,b]}\|f^{(2Q)} (A,x)\|,\]
	where the subscript $(q)$ refers to the $q$th-order partial derivative with respect to $x$.
\end{lemma}
\begin{proof}
	Since $A$ is a Hermitian matrix, there exists a unitary matrix $U$ such that $A = U\Lambda U^{-1}$, where $\Lambda = \text{diag}(\lambda_1,\cdots,\lambda_n)$, which yields
	\[\Big\|\int_a^b f(A,x) \d x - \sum_{q=0}^{Q-1} w_q f(A,x_q) \Big\| = \Big\|\int_a^b f(\Lambda,x) \d x - \sum_{q=0}^{Q-1} w_q f(\Lambda,x_q) \Big\|.\]
	According to Chapter 5 of \cite{KMN1989}, there exists $\xi_j \in (a,b)$ such that
	\[ \int_a^b f(\lambda_j, x) \d x - \sum_{q=0}^{Q-1}w_q f(\lambda_j, x_q)  = \frac{(b-a)^{2Q+1}(Q!)^4 }{(2Q+1)[ (2Q)! ]^3} f^{(2Q)} (\lambda_j,\xi_j).\]
	Therefore,
	\[ \int_a^b f(\Lambda,x) \d x - \sum_{q=0}^{Q-1} w_q f(\Lambda,x_q)
	= \frac{(b-a)^{2Q+1}(Q!)^4 }{(2Q+1)[ (2Q)! ]^3} \begin{bmatrix}
		f^{(2Q)} (\lambda_1,\xi_1)  &    &    \\
		&   \ddots  &  \\
		&           & f^{(2Q)} (\lambda_n,\xi_n)
	\end{bmatrix},\]
	which implies the desired estimate.
\end{proof}

\begin{lemma}\label{lem:error1}
	Let $p(A)$ be defined in \eqref{ShrodingerhA}. Under the condition of Theorem \ref{thm:pdiscretization},  the quadrature error can be bounded as
		\begin{equation}\label{suberror}
			\|g(A) -  p(A)\| \le \delta_1
		\end{equation}
	when we choose
	\[\tau = \frac{1}{\|D_\mu \otimes A\|}, \quad Q = \Theta\Big( \log \frac{ T }{\delta_1} \Big).\]
\end{lemma}
\begin{proof}
	The error can be decomposed as
	\[ \|g(A) -  p(A)\| \le \Big\| g(A) - \sum_{m=0}^{N_t-1} \sum_{q=0}^{Q-1} c_{m,q} \e^{\i (D_\mu \otimes A)t_{m,q}}\Big\| =: I. \]
By Lemma \ref{lem:ShrodingerGaussErr},
	\begin{align*}
	\Big\|  \int_0^{T} U(A,t) \d t - \sum_{m=0}^{N_t-1} \sum_{q=0}^{Q-1}  c_{m,q} U(A,t_{m,q})\Big\|
		 \le  \frac{N_t \tau^{2Q+1}(Q!)^4 }{(2Q+1)[ (2Q)! ]^3} \max_t \|U^{(2Q)}(A,t) \|.
	\end{align*}
It is simple to find that
	\[\max_t \|U^{(2Q)}(A,t) \| \le \|D_\mu \otimes A\|^{2Q},\]
which gives
	\[I \le T \frac{\tau^{2Q}(Q!)^4 }{(2Q+1)[ (2Q)! ]^3} \|D_\mu \otimes A\|^{2Q}.\]
	If we choose $\tau = \frac{1}{\|D_\mu \otimes A\|}$, then
	\[I \le \frac{T(Q!)^4 }{(2Q+1)[ (2Q)! ]^3}. \]
	Therefore, we obtain $I \le \delta$ by choosing $Q$ such that
	\[\frac{(Q!)^4 }{(2Q+1)[ (2Q)! ]^3}   \le \frac{\delta_1}{T}.\]
	As before, it suffices to choose $Q = \Theta( \log \frac{ T }{\delta_1} )$, which completes the proof.
\end{proof}

Combining the result in Theorem \ref{thm:pdiscretization}, we are ready to obtain the approximation error.

\begin{theorem}\label{thm:ShrodingerhAErr}
Let $A$ be an invertible Hermitian matrix. Suppose that $\bb{x}$ is the exact solution of $A\bb{x} = \bb{b}$ and $\bb{x}_{h,T}^{\text{d}}$ is the numerical solution defined in \eqref{xTtruncateGauss}, respectively. Under the condition of Theorem \ref{thm:pdiscretization}, if we choose
\[\delta_1 \simeq \frac{\delta \xi}{\|\bb{\psi}\|}, \qquad
\delta_2 \simeq  \frac{\delta_1}{T}, \qquad T = \Theta\Big(\frac{\kappa}{\|A\|}  \sqrt{2 \log\frac{1}{\delta}}\Big),
\]
\[ \tau = \frac{T}{N_t} = \frac{1}{\|D_\mu \otimes A\|}, \qquad Q = \mathcal{O}\Big( \log \frac{\kappa}{\xi \|A\|\delta } \Big),\]
then there holds
\[ \| \bb{x}- \bb{x}_{h,T}^{\text{d}} \| \lesssim \delta \|\bb{x}\| + \delta \xi \|\bb{b}\|. \]
\end{theorem}
\begin{proof}
By definition,
\begin{align*}
\| p(A) - \tilde{p}(A)\|
& = \Big\| \sum_{m=0}^{N_t-1} \sum_{q=0}^{Q-1} c_{m,q} (\e^{\i (D_\mu \otimes A)t_{m,q}} - V_{m,q})\Big\| \\
& \le \sum_{m=0}^{N_t-1} \sum_{q=0}^{Q-1} |c_{m,q}|  \delta_2 \lesssim T \delta_2.
\end{align*}
According to Eq.~\eqref{suberror} in Lemma \ref{lem:error1}, we obtain from $\|\tilde{\bb{W}}_h(0)\| = \|\bb{W}_h(0)\|$ $ = \|\bb{\psi}\| \|\bb{b}\|$ that
\begin{align}
\|\bb{x}_{h,T}-\bb{x}_{h,T}^{\text{d}}\|
& \le \| g(A) - \tilde{p}(A)\| \|\tilde{\bb{W}}_h(0)\| \nonumber\\
&  (\| g(A) - p(A)\| + \| p(A) - \tilde{p}(A)\|) \|\bb{W}_h(0)\| \nonumber\\
& \le (\delta_1 + T \delta_2) \|\bb{\psi}\| \|\bb{b}\| \simeq \delta \xi \|\bb{b}\|,  \label{errgA}
\end{align}
provided that
\[\delta_1 \simeq \frac{\delta \xi}{\|\bb{\psi}\|}, \qquad
\delta_2 \simeq \frac{\delta \xi }{T\|\bb{\psi}\|} = \frac{\delta_1}{T}.
\]
This yields
\[
\log \frac{T}{\delta_1} \simeq \log \Big(\frac{\kappa\|\bb{\psi}\|}{\xi \|A\|\delta } \log^{1/2}\frac{1}{\delta} \Big), \qquad
\log \frac{1}{\delta_2} \simeq \log \frac{T}{\delta_1} .
\]
Noting that
\[
\|\bb{\psi}\| \le \Big(\frac{1}{\Delta p} \int_{\mathbb{R}} |\psi(p)|^2 \d p\Big)^{1/2}
\lesssim \Big(\frac{1}{\Delta p} \Big)^{1/2} \lesssim \log^{3/2} \frac{\kappa }{\xi \|A\| \delta}
\]
and $\frac{\kappa }{\xi \|A\|} \ge 1$, we then obtain
\[
\log \frac{T}{\delta_1}\simeq \log \frac{1}{\delta_2} \lesssim \log \Big(\frac{\kappa}{\xi \|A\|\delta }\log^2 \frac{\kappa }{\xi \|A\| \delta} \Big)
\lesssim \log \frac{\kappa}{\xi \|A\|\delta }.\]
The proof is completed by combining \eqref{errgA} and \eqref{error2}.
\end{proof}

\subsection{Linear combination of unitaries}\label{subsec:complexity_Sch}

For simplicity we rewrite the summation in \eqref{ShrodingerhA} by a single index as
\begin{equation}
p(A) = \sum_{m=0}^{N_t-1} \sum_{q=0}^{Q-1} c_{m,q} \e^{\i (D_\mu \otimes A)t_{m,q}} =: \sum_{j=0}^{M-1} \alpha_j U_j,
\end{equation}
where
\[U_j = \e^{\i (D_\mu \otimes A)t_j} = (\e^{\i (D_\mu \otimes A)\tau})^j.\]
It is evident that $M \le N_t Q$ since we have collected the like terms. Moreover, the initial data $\tilde{\bb{w}}_h(0)$ is given in \eqref{discreteLCHSQLSP}.

In the following we will present the details on how to apply the LCU technique to effectively prepare the solution state for $A \bb{x} = \bb{b}$. To this end, we need to pre-construct the following oracles.
\begin{itemize}
	\item The coefficient oracles
	\[O_{\text{coef}}: \ket{0^{n_a}} \to \frac{1}{\sqrt{\|\bb\alpha\|_1}} \sum\limits_{j=0}^{M-1}\sqrt{\alpha_j} \ket{j}, \qquad M = 2^{n_a},\]
	
	with  $\bb\alpha = (\alpha_0, \cdots, \alpha_{M-1})$ and $\|\bb\alpha\|_1 = \alpha_0 + \cdots + \alpha_{M-1}$.
	It is obvious that
	\[\|\bb\alpha\|_1 = \sum_{m=0}^{N_t-1} \sum_{q=0}^{Q-1} |c_{m,q}|
	= \mathcal{O}(T),\]
	where the last equality is obtained from $\sum_{q,m} |c_{q,m}| = \int_0^T  \d t = T$.
	
	\item The select oracle
	\[\text{SEL}_A = \sum_{j=0}^{M-1} \ket{j}\bra{j} \otimes U_j, \qquad U_j = (\e^{\i (D_\mu \otimes A)\tau})^j.\]

	\item The state preparation oracle
	\[O_w: \ket{0^{n_w}} \to \ket{\tilde{W}_h(0)},\]
where $\ket{\tilde{W}_h(0)}$ is the quantum state for $\tilde{\bb{W}}_h(0)$.
\end{itemize}

\begin{theorem}\label{lem:time_complexity_Sch}
Let $A$ be an invertible Hermitian matrix and assume that $\bb{x}$ is the solution to $A \bb{x} = \bb{b}$.  Then there exists a quantum algorithm that prepares an $\mathcal{O}(\delta)$-approximation of the state $\ket{x} = \bb{x}/\|\bb{x}\|$ with $\Omega(1)$ success probability and a flag indicating success, using
\[\mathcal{O}\Big(  \frac{T \|\bb{\psi}\|}{\xi} \Big)\]
 queries to the coefficient oracle $O_{\text{coef}}$, the select oracle $\text{SEL}_A$ and the state preparation oracle $O_w$, where
\[T =  \mathcal{O}\Big(\frac{\kappa}{\|A\|} \sqrt{2\log\frac{1}{\delta}}\Big), \qquad
  \|\bb{\psi}\| = \mathcal{O}\Big(\log^{3/2} \frac{\kappa }{\xi \|A\| \delta}\Big), \qquad \xi = \|A^{-1}\ket{b} \|.\]
\end{theorem}
\begin{proof}
The procedure of preparing an approximation of the state proportional to
\[\bb{W}_{\text{int}}:=\int_0^T \bb{W}_h(t) \d t  = \sum_{k = 0}^{N_p-1} \ket{k} \otimes \int_0^T \bb{w}_h(t,p_k)  \d t = (\Phi \otimes I)g(A) \tilde{\bb{W}}_h(0) \]
is given as follows:
	\begin{align*}
		\ket{0^{n_a}} \otimes \ket{0^{n_w}}
		&\quad \xrightarrow{O_{\text{coef}}~\otimes~ I^{\otimes n_w}} \quad \frac{1}{\sqrt{\|\bb\alpha\|_1}} \sum_{j=0}^{M-1}\sqrt{\alpha_j} \ket{j} \otimes \ket{0^{n_w}} \\
		&\quad \xrightarrow{~I^{\otimes n_a}~ \otimes ~O_w~~} \quad \frac{1}{\sqrt{\|\bb\alpha\|_1}} \sum_{j=0}^{M-1}\sqrt{\alpha_j} \ket{j} \otimes \ket{\tilde{W}_h(0)}\\
		&\quad \xrightarrow{~~~~~\text{SEL}_A~~~~~~} \quad \frac{1}{\sqrt{\|\bb\alpha\|_1}} \sum_{j=0}^{M-1}\sqrt{\alpha_j} \ket{j} \otimes U_j \ket{\tilde{{W}}_h(0)}\\
		&\quad \xrightarrow{O_{\text{coef}}^\dag~\otimes~ I^{\otimes n_w}} \quad \frac{1}{\|\bb\alpha\|_1} \ket{0^{n_a}} \otimes \sum_{j=0}^{M-1}\alpha_j U_j \ket{\tilde{{W}}_h(0)}+\ket{\bot} \\
&\quad \xrightarrow{~I^{\otimes n_a}~\otimes~\Phi~\otimes~I^{\otimes n}} \quad \frac{1}{\|\bb\alpha\|_1} \ket{0^{n_a}} \otimes (\Phi \otimes I) p(A) \ket{\tilde{{W}}_h(0)}+\ket{\bot'}.
	\end{align*}
Denote by $V$ the resulting unitary operator. Then,
	\begin{align*}
		\ket{0^{n_a}} \otimes \ket{0^{n_w}}
		\quad
		\xrightarrow{~V~} &\quad \frac{1}{\|\bb\alpha\|_1} \ket{0^{n_a}} \otimes (\Phi \otimes I) p(A) \ket{\tilde{{W}}_h(0)}+\ket{\bot'}\\
		=&\quad \frac{1}{\|\bb\alpha\|_1 \|\tilde{\bb{W}}_h(0)\|} \ket{0^{n_a}} \otimes \bb{W}_{\text{int}}^{\text{d}} +\ket{\bot'},
	\end{align*}
where $\bb{W}_{\text{int}}^{\text{d}}$ is the approximation of $\bb{W}_{\text{int}}$ obtained through numerical integration.

The final approximate state $\ket{x_{h,T}^{\text{d}}}$ corresponding to $\bb{x}_{h,T}^{\text{d}} = \Pi_*\bb{W}_{\text{int}}^{\text{d}}$ is obtained by projection.
Using the inequality $\| \frac{\bb{x}}{\|\bb{x}\|} - \frac{\bb{y}}{\|\bb{y}\|} \| \le 2 \frac{\|\bb{x} - \bb{y}\|}{\|\bb{x}\|}$ for two vectors $ \bb{x}, \bb{y} $ and using the estimate in Theorem \ref{thm:ShrodingerhAErr}, we can bound the error in the quantum state after a successful measurement as
\begin{align*}
\|\ket{x} - \ket{x_{h,T}^{\text{d}}}\|
 \le 2 \frac{\|\bb{x} - \bb{x}_{h,T}^{\text{d}}\|}{\|\bb{x}\|}
\lesssim  \delta +  \delta \xi \frac{ \|\bb{b}\|}{\|\bb{x}\|} = 2\delta.
\end{align*}

The state $\ket{W_{\text{int}}^{\text{d}}}$ is obtained by measuring the state and obtaining all 0 in the other qubits. The likelihood of acquiring this approximate state is
\[\text{P}_w= \Big(\frac{\|\bb{W}_{\text{int}}^{\text{d}}\|}{\|\bb\alpha\|_1\|\tilde{\bb{W}}_h(0)\|}\Big)^2.\]
To get the final approximate state $\ket{x_{h,T}^{\text{d}}}$, we need to perform the projection onto $\ket{k_*}$ as shown in \eqref{xT}, with the probability given by
\[\text{P}_x= \Big(\frac{\|\bb{w}_{*}^{\text{d}}\|}{\|\bb{W}_{\text{int}}^{\text{d}}\|}\Big)^2,\]
where $\bb{w}_{*}^{\text{d}}$ is the numerical approximation of $\int_0^T \bb{W}_h(t,0) \d t =\bb{x}_{h,T} \approx \bb{x}$. The overall probability is
\[\text{P}_r = \Big(\frac{\|\bb{w}_{*}^{\text{d}}\|}{\|\bb\alpha\|_1\|\bb{W}_h(0)\|}\Big)^2.\]
The success probability can be raised to $\Omega(1)$ by using $\mathcal{O}(g)$ rounds of amplitude amplification, where
\begin{align*}
g  = \frac{\|\bb\alpha\|_1\|\bb{W}_h(0)\|}{\|\bb{w}_{*}^{\text{d}}\|}
= \frac{\|\bb\alpha\|_1 \|\bb{\psi}\| \|\bb{b}\|}{\|\bb{w}_{*}^{\text{d}}\|}
\approx \frac{\|\bb\alpha\|_1 \|\bb{\psi}\| \|\bb{b}\|}{\|\bb{x}\|}
= \mathcal{O}\Big( \frac{T \|\bb{\psi}\|}{\|A^{-1} \ket{b}\|} \Big).
\end{align*}
The proof is finished by combining the result in Theorem \ref{thm:ShrodingerhAErr}.
\end{proof}

Thus, our algorithm for query complexity is consistent with that presented in \cite{Childs2017QLSA}, implying that the final time complexity will exhibit quadratic dependence on the condition number $ \kappa $ if we include the cost of the Hamiltonian simulation $U(A,t)=\e^{\i (D_\mu \otimes A)t}$.
\begin{theorem}\label{lem:queries_complexity_Sch}
Let $ A $ be an invertible matrix, and consider the linear system $ A \bb{x} = \bb{b} $. Suppose that we are given the block encoding of $A$ with normalization factor $\alpha_A \ge \|A\|$ and that an upper bound on its inverse $\alpha_{A^{-1}} \ge \|A^{-1}\|$ is known. Let $\kappa_A = \alpha_A \alpha_{A^{-1}}$ be an upper bound on the condition number $\kappa = \|A\| \|A^{-1}\|$ of $A$. Then, there exists a quantum algorithm that prepares an $\mathcal{O}(\delta)$-approximation of the state $\ket{x}$ with $\Omega(1)$ success probability and a flag indicating success, using
\begin{equation}\label{log15}
\mathcal{O}\Big(\frac{\kappa_A^2}{\xi \alpha_A} \log^{7.5} \frac{\kappa_A}{\xi \alpha_A \delta}\Big), \qquad
	\xi = \|A^{-1}\ket{b} \|,
\end{equation}
    queries to the block-encoding oracle for $A$ and
\[\mathcal{O}\Big(\frac{\kappa_A}{\xi \alpha_A} \log^2 \frac{\kappa_A}{\xi \alpha_A \delta}\Big)\]
    queries to the state preparation oracle for $\bb{b}$.
\end{theorem}
\begin{proof}
(1) Without loss of generality, we can assume that $A$ is a Hermitian matrix. According to Theorem \ref{lem:time_complexity_Sch}, we invoke the select oracle and the state preparation oracle $\mathcal{O}( g )$ times with
\begin{equation}\label{costb}
g \lesssim  \frac{T \|\bb{\psi}\|}{\xi} \lesssim \frac{\kappa}{\xi\|A\|}\log^2 \frac{\kappa }{\xi \|A\| \delta}.
\end{equation}

The query complexity of simulating a Hamiltonian $U(A,t)=\e^{\i (D_\mu \otimes A)\tau}$ with error at most $\delta_2$ is
\[C_\tau = \mathcal{O}\Big( \mu_{\max} \alpha_A  \tau + \log \frac{1}{\delta_2} \Big), \qquad \mu_{\max} =  \mathcal{O}\Big( \log^3 \frac{\kappa}{\xi \|A\| \delta} \Big).\]
The select oracle
\[\text{SEL}_A = \sum_{j=0}^{M-1} \ket{j}\bra{j} \otimes U_j, \qquad U_j = (\e^{\i (D_\mu \otimes A)\tau})^j\]
can be implemented with query complexity
\begin{align*}
C_{\text{SEL}}
& \lesssim M C_\tau \lesssim Q \Big( \mu_{\max} \alpha_A  T +  N_t \log \frac{1}{\delta_2} \Big) \\
& \lesssim Q \Big( \mu_{\max} \alpha_A  T +  \mu_{\max}\alpha_A T \log \frac{1}{\delta_2} \Big) \\
& \lesssim Q\mu_{\max} \alpha_A  T  \log \frac{1}{\delta_2} \lesssim \frac{\alpha_A}{\|A\|}\kappa \log^{5.5} \frac{\kappa }{\xi\|A\| \delta}.
\end{align*}
Therefore, we use
\begin{equation}\label{costA}
\mathcal{O}\Big( \frac{\alpha_A}{\|A\|} \frac{\kappa^2}{\xi \|A\|} \log^{7.5} \frac{\kappa}{\xi\|A\| \delta}\Big), \qquad
	\xi = \|A^{-1}\ket{b} \|,
\end{equation}
 queries to the block-encoding oracle for $A$.

(2) It is evident that
\[\kappa = \|A\| \|A^{-1}\| \le \alpha_A \alpha_{A^{-1}} = \kappa_A,\]
\[\frac{\kappa}{\|A\|} =  \|A^{-1}\| \le \alpha_{A^{-1}} = \frac{\alpha_A\alpha_{A^{-1}}}{\alpha_A} = \frac{\kappa_A}{\alpha_A}.\]
These inequalities allow us to replace $\kappa$ and $\kappa/\|A\|$ in \eqref{costA} and \eqref{costb} by $\kappa_A$ and $\kappa_A/\alpha_A$, respectively, thereby eliminating the factor $\alpha_A/\|A\|$ in \eqref{costA}. The proof is completed.
\end{proof}

%\begin{remark}\label{rem:quadratic}
%The factor $\xi \|A\|$ is invariant under the transformation $A\to c A$, where $c$ is a constant. In the best case scenario that $\ket{b}$ has an $\Omega(1)$ overlap with the left-singular vector of $A$ with respect to the smallest singular value,  $\xi \|A\| = \|A^{-1} \ket{b}\| \|A\| = \Omega(\kappa)$ and hence the nearly linear run time $\mathcal{O}(\kappa \text{polylog}(\kappa/\delta))$ is obtained with respect to the condition number $\kappa$.
%\end{remark}

\section{Improved dependence on condition number via preconditioning} \label{sec:preconditioning}

In this section, we demonstrate that the block preconditioning technique introduced in \cite{Low2026quantumlinearsystem} can be applied to achieve nearly linear scaling in the condition number $\kappa_A$.

Let $\Pi_{\text{b}} = \ket{b}\bra{b}$ be the projection operator onto $\ket{b}$. For any constant $0<s<1$, we define $S = I - (1-s)\Pi_{\text{b}}$, which is referred to as the block preconditioner in \cite{Low2026quantumlinearsystem}, and consider the following preconditioned linear system
\begin{equation}\label{SASb}
SA \bb{x} = S \bb{b}.
\end{equation}
For the quantum computation, we should assume the state preparation oracle
\[O_{Sb}: \ket{0^n} \to  \ket{S b} := \frac{S\bb{b}}{\|S\bb{b}\|}.\]
However, since $\ket{S b} = \ket{b}$, we can take $O_{Sb} = O_b$, where $O_b$ is the state preparation oracle for $\bb{b}$. According to Eq.~(176) in \cite{Low2026quantumlinearsystem}, we can block encode $S$ with normalization factor 1 using two queries to $O_b$. This yields the block-encoding oracle for $SA$ with normalization factor $\alpha_A$.

For the preconditioned system \eqref{SASb}, we summarize its properties below with the details provided in \cite{Low2026quantumlinearsystem}.
\begin{lemma} \label{lem:Sproperty}
Let $A$ be an invertible matrix. Suppose we have a constant multiplicative approximation of the solution norm $\xi = \|A^{-1}\ket{b}\|$, denoted by $\xi_c$, i.e., there exists a constant $c>1$ such that
\[\frac{\xi}{c} < \xi_c < c \xi.\]
Let $\bb{x} = A^{-1} \ket{b}$ and $\bb{y} = (SA)^{-1} \ket{Sb}$. If we choose $s = \frac{\xi_c}{c\alpha_{A^{-1}}}$, which satisfies
\begin{equation}\label{sChoose}
\frac{\xi}{c^2\alpha_{A^{-1}}} < s < \frac{\xi}{\alpha_{A^{-1}}} \le \frac{\xi}{\|A^{-1}\|} \le 1,
\end{equation}
then we have
\begin{align}
& \xi_{SA} = \|\bm{y}\| = \frac{1}{s} \|\bb{x}\| = \frac{1}{s} \xi, \label{property1}\\
& \|SA\| \le \|A\| \le \alpha_A,\label{property2}\\
&\|(SA)^{-1}\| \le \sqrt{c^4 + 1} \alpha_{A^{-1}}.\label{property3}
\end{align}
\end{lemma}

Combining the above discussion, we obtain the nearly linear dependence.

\begin{theorem}\label{the:finall_op_complexty}
Let $A$ be an invertible matrix and consider the preconditioned linear system \eqref{SASb}. Under the conditions of Lemma \ref{lem:Sproperty}, there exists a quantum algorithm that prepares an $\mathcal{O}(\delta)$-approximation of the state $\ket{x} = \ket{y}$ with $\Omega(1)$ success probability and a flag indicating success, using
\begin{equation}\label{oplog75}
\mathcal{O}\Big(\kappa_A \log^{7.5} \frac{1}{\delta}\Big),
\end{equation}
queries to the block-encoding oracle for $A$ and the state preparation oracle for $\bb{b}$.
\end{theorem}
\begin{proof}
According to Theorem \ref{lem:queries_complexity_Sch}, the linear system \eqref{SASb} can be solved with
\begin{equation}\label{cpSA}
g = \mathcal{O}\Big(\frac{\kappa_{SA}^2}{\xi_{SA} \alpha_{SA}} \log^{7.5} \frac{\kappa_{SA}}{\xi_{SA} \alpha_{SA} \delta}\Big), \qquad
	\xi_{SA} = \|(SA)^{-1}\ket{Sb} \|,
\end{equation}
queries to the block-encoding oracle for $SA$ and
\[\mathcal{O}\Big(\frac{\kappa_{SA}}{\xi \alpha_{SA}} \log^2 \frac{\kappa_{SA}}{\xi_{SA} \alpha_{SA} \delta}\Big)\]
queries to the state preparation oracle for $\ket{Sb} = \ket{b}$. Since block encoding $S$ uses two queries to $O_b$, for this preconditioned system, the query complexity for $A$ and $\bb{b}$ are comparable. According to Lemma \ref{lem:Sproperty} and the definition of $\alpha_A$, we can take
\[\alpha_{SA} = \alpha_A, \qquad \alpha_{(SA)^{-1}} = \sqrt{c^4 + 1} \alpha_{A^{-1}}.\]
This gives
\[\kappa_{SA} = \alpha_{SA} \alpha_{(SA)^{-1}} \lesssim \alpha_A\alpha_{A^{-1}} = \kappa_A.\]
Combining with \eqref{property1}, we can rewrite \eqref{cpSA} as
\[
g = \mathcal{O}\Big(\frac{s\kappa_A^2}{\xi \alpha_A} \log^{7.5} \frac{s\kappa_A}{\xi \alpha_A \delta}\Big), \qquad
	\xi  = \|A^{-1}\ket{b} \|.
\]
This together with the choice of $s$ in \eqref{sChoose} yields
\[\frac{s\kappa_A}{\xi \alpha_A} = \frac{s\alpha_A\alpha_{A^{-1}}}{\xi \alpha_A} < 1. \]
The proof is completed.
\end{proof}

%\begin{remark}
	%From Theorem~18 of \cite{Costa2022QLSA} or Section~5 of \cite{dalzell2026shortcutoptimalquantumlinear}, the cost of estimating $\xi_c$ scales linearly with the condition number $\kappa$ of the coefficient matrix, i.e., $\mathcal{O}(\kappa)$. 
	%Therefore, an additional factor of $\mathcal{O}(\kappa)$ is introduced into our complexity. 
	%This factor can be absorbed into the overall complexity, so the algorithm still achieves optimal performance.
%\end{remark}

%\begin{remark}
%	In fact, we note that if we take $s = \frac{\xi}{\|A^{-1}\|} = 1$, then $\bb{b}$ is exactly the minimal left singular vector of $A$. In this case, no preconditioning is needed, and the linear system solver still achieves near-optimal query complexity $\mathcal{O}\left(\kappa_{A} \log^{7.5} (1/\delta)\right)$.
%\end{remark}

\section*{Acknowledgments}

Y Yang was supported by NSFC grant No.\ 12571469, the Project of Scientific Research Fund of the Hunan Provincial Science and Technology Department (No.\ 2024JJ1008), the Major Scientific and Technological Innovation Platform Project of Hunan Province (2024JC1003), the 111 Project (No.\ D23017), and Program for Science and Technology Innovative Research Team in Higher Educational Institutions of Hunan Province of China.
Y Yu was supported by NSFC grant (No.\ 12301561), the Key Project of Scientific Research Project of Hunan Provincial Department of Education (No.\ 24A0100), the Science and Technology Innovation Program of Hunan Province (No.\ 2025RC3150) and the general program of Hunan Provincial Natural Science Foundation (No.\ 2026JJ50003).
L Zhang was supported by the Hunan Provincial Graduate Student Research and Innovation Project (No.\ CX20250933) and the Xiangtan University Graduate Student Research and Innovation Project (No.\ XDCX2025Y188).
This research was supported in part by the 111 Project (No.\ D23017), and Program for Science and Technology Innovative Research Team in Higher Educational Institutions of Hunan Province of China.

\bibliographystyle{plain} %plain, unsrt, alpha
\bibliography{Refs}

\end{document}